 \documentstyle[prl,aps]{revtex}  
\begin{document}  
\draft  
\preprint{\today}  
  
  
 \twocolumn[\hsize\textwidth\columnwidth\hsize  
 \csname @twocolumnfalse\endcsname              
  
\title{N=Z Waiting Point Nucleus $^{68}$Se:  
       Collective Rotation and High-K Isomeric States}  
\author{Yang Sun$^{1}$, Zhanwen Ma$^{1}$,   
        Ani Aprahamian$^{2}$, Michael Wiescher$^{2}$}  
\address{  
$^1$Department of Physics and Astronomy, University of Tennessee,  
Knoxville, Tennessee 37996\\  
$^2$Department of Physics, University of Notre Dame,  
Notre Dame, Indiana 46556\\  
}  
  
\date{\today}  
\maketitle  
  
\begin{abstract}  
The structure of collective states  
and quasi-particle excitations in $^{68}$Se  
is analyzed by using the projected shell model.  
It is found that the   
$g_{9/2}$ orbitals play a decisive role for the structure of this nucleus.   
At the prolate minimum, alignment of the low-K states causes   
the observed first backbending in moment of inertia    
and we predict a second backbending that has not yet been reported.   
At the oblate minimum, quasi-particles having    
high-K can form long-lived isomeric states.   
The structure of $^{68}$Se is discussed with respect to the isomers  
and their possible impact on the determination of its nuclear mass.  
\end{abstract}  
  
  
 ]  
  
\narrowtext  
\bigskip  
Thermonuclear runaways on the surface of neutron stars have been identified 
as energy sources for type I X-ray bursts 
\cite {Woos76,Maras77,Joss77}. The explosive conditions resulting  
from the thermonuclear runaway provide the site for rapid sequences of successive proton captures (rp-process) resulting in the creation of nuclei far  
beyond $^{56}$Ni all the way to the proton rich regions \cite{VanWorm}  
of the chart of nuclides.  
  
X-ray bursts are observed frequently \cite{PhysicsToday},  
yet the nucleosynthesis and the correlated energy generation  
is not completely understood. In addition to the uncertainty of  
the astrophysical conditions, network simulations of the rp-process  
are hindered by the lack of experimental information on the structure  
of nuclei along the rp-process path \cite{VanWorm,SchatzRep}.  
Nuclei of particular interest to the rp-process are the N=Z  
waiting point nuclei \cite{SchatzRep}.  
These are the nuclei where further p-capture  
typically leads to a p-unbound isotone  
or where the Q value for the (p,$\gamma$) reaction is low making  
the reverse  
reaction possible. In order for the rp-process to continue,,  
it needs to wait for $\beta$-decay or for the 2p-capture  
to bridge the waiting point.  
The time spent at these N=Z waiting points  
is strongly dependent on the photodisintegration rates,  
the $\beta$-decay lifetimes, and particularly on the masses  
of nuclei along the path.   
  
The mass of $^{68}$Se is particularly relevant to determining  
whether further p-capture is likely to occur.  
There have been several attempts \cite{mass}  
to measure the half-life of $^{69}$Br by fragmentation studies  
at GANIL and MSU. No $^{69}$Br was observed but an upper  
limit of 150ns was set for the half-life. The implication is  
that $^{69}$Br should be p-unbound. This implies that  
the rp-process should proceed via 2p-capture or $\beta$-decay of $^{68}$Se.   
There are some new mass   
measurements \cite{Lalle2000} of several N=Z nuclei in the A=80 mass  
region from GANIL. The measurement involves the use of the CSS2  
cyclotron at GANIL as a high resolution mass spectrometer.  
The relative mass differences of unknown and known nuclei  
are determined from their  
differences in times of flight in the CSS2.  
The mass excess measured for $^{68}$Se was $-52.347\pm$0.080 MeV.  
If one now uses the measured $^{68}$Se mass, the implication is  
that $^{69}$Br should be bound by greater than 1 MeV.  
This result is of course inconsistent with the fragmentation  
experiments. One possible explanation for the discrepancy  
in the measurements would be if the GANIL mass measurement was  
that of an isomeric state in $^{68}$Se.  
  
The purpose of this Letter is to investigate the role  
of the structure of $^{68}$Se for the mass measurements  
and therefore for the rp-process. We analyze the structure of  
both collective and quasi-particle excitations in $^{68}$Se  
using the projected shell model (PSM) \cite {review} with the idea  
of looking for the feasibility of the existence of isomeric  
states in this nucleus.  
  
Great progress has been made on the theory side by the application  
of advanced shell model diagonalization methods  
which can explicitly yield the 
structure wave functions, and matrix elements for  
even-even, odd-A, and odd-odd nuclei.  
One successful example is the application of the   
spherical pf-shell model developed by the  
Strasbourg-Madrid group \cite{SM}.   
By employing this model,   
Langanke and Mart\'\i nez-Pinedo have demonstrated its strength   
in understanding the nuclear processes that govern those violent astrophysical  
phenomena \cite{La00}.   
  
The rp-process path involves nuclei which are significantly deformed.  
At the deformed potential minimum, abnormal high-$j$ orbitals such as   
$g_{9/2}$   
intrude into the pf-shell near the Fermi level.   
Therefore, the $g_{9/2}$ orbitals have to be included in the model 
space for the present interest.  
This strongly suggests proper selection of a shell model basis that   
is capable of  
describing the undergoing physics within a manageable space.   
 
The PSM \cite{review} 
is a shell model diagonalization method which  
adopts the bases from the well-established Nilsson model   
with pairing correlation incorporated into them by a BCS  
calculation. The selection of the bases is first implemented in the  
multi-quasiparticle (qp) basis with respect to the deformed BCS vacuum  
$\left|0\right>$; then the broken rotational  
symmetry is recovered by  
angular momentum projection technique \cite{review} to  
form a shell model basis in the laboratory frame.  
Finally a shell model Hamiltonian is  
diagonalized in the projected space.  
The problem of dimensionality that is often  
tied with conventional shell-model calculations does not occur in the PSM.   
   
In the present PSM calculation for $^{68}$Se, particles in    
three major shells ($N=2,3,4$) for both neutron and proton are  
activated.   
The shell model space includes the   
0-, 2- and 4-qp states:  
\begin{equation}  
\left\{\left|0 \right>, \   
\alpha^\dagger_{n_i} \alpha^\dagger_{n_j} \left|0 \right>,\  
\alpha^\dagger_{p_i} \alpha^\dagger_{p_j} \left|0 \right>,\  
\alpha^\dagger_{n_i} \alpha^\dagger_{n_j} \alpha^\dagger_{p_i}  
\alpha^\dagger_{p_j} \left|0 \right> \right\} ,  
\label{baset}  
\end{equation}  
where $\alpha^\dagger$ is the creation operator for a qp and the  
index $n$ ($p$) denotes neutron (proton) Nilsson quantum numbers which run over  
properly selected (low-lying) orbitals. 
Note that the index $i$ and $j$ in Eq. (\ref{baset}) are general. For  
example, a 2-qp state can be of positive parity if both quasiparticles  
$i$ and $j$ are from the same major shell; it can also be of negative  
parity if two quasiparticles are from two neighboring major shells.  
Positive and negative parity states span the whole configuration space  
with the corresponding matrix in a block-diagonal form classified by the  
parity.  
  
As the first step in a PSM calculation,  
one has to find out where the optimal basis is.   
The shape coexistence feature in $^{68}$Se seems to be firmly suggested   
by the recent experimental data of Fischer {\it et al.} \cite{Fi00}.  
Two coexisting rotational bands were identified, with the ground state band  
having properties consistent with collective oblate rotation, and the   
excited band having characteristics consistent with prolate rotation.  
Constrained Hartree-Fock calculations by Sarriguren {\it et al.} \cite{Moya99}   
also found coexistence of two energy minima in $^{68}$Se,   
with the oblate solution  
as the ground state, in supporting the experimental conclusion.   
Moreover,   
the observed bands are found to interact rather weakly  
indicating a high energy barrier \cite{Fi00}.   
The theoretical barrier between the two well-separated   
minima is calculated as about 3 MeV \cite{Moya99}.   
These evidences suggest a possible simplification in our calculation    
that we can perform the diagonalization separately at the prolate  
and oblate minimum by neglecting the coupling between the two minima.     
Collecting all of the above information, we thus   
perform the following PSM   
calculations at $\varepsilon_2=0.28$ for the prolate states,  
and $\varepsilon_2=-0.24$ for the oblate states.   
  
It is evident from the experimental data \cite{Fi00}   
that the rotational states  
in $^{68}$Se are dominated by quadrupole collectivity and  
pairing interactions. To describe these properties, it is  
efficient to use a quadrupole plus pairing Hamiltonian \cite{review}    
\begin{equation}  
\hat H = \hat H_0 - {1 \over 2} \chi \sum_\mu \hat Q^\dagger_\mu  
\hat Q^{}_\mu - G_M \hat P^\dagger \hat P - G_Q \sum_\mu \hat  
P^\dagger_\mu\hat P^{}_\mu,  
\label{hamham}  
\end{equation}  
where $\hat H_0$ is the spherical single-particle Hamiltonian which  
contains a proper spin-orbit force, whose strengths (i.e.  
the Nilsson parameters) are taken from Ref.  
\cite{Sun00}.   
The second term in the Hamiltonian is the   
quadrupole-quadrupole interaction and the last  
two terms are the monopole and quadrupole pairing interactions,  
respectively.   
The interaction strength $\chi$ is determined by the self-consistent  
relation such that the input quadrupole deformation $\varepsilon_2$ and  
the one resulting from the HFB procedure coincide with each other  
\cite{review}. The monopole pairing strength $G_M$ is taken to be  
$G_M=\left[18.0-14.5(N-Z)/A\right]/A$ for neutrons and $G_M=14.5/A$ for  
protons.   
Finally, the quadrupole pairing  
strength $G_Q$ is assumed to be proportional to $G_M$, the  
proportionality constant  
being fixed to 0.20 in the present work.  
These interaction strengths are consistent   
with the values used in the previous PSM calculations  
for this mass region \cite{Do98,Pa00}.  
  
The eigenvalue equation of the PSM for a given spin $I$ takes the  
form \cite{review}  
\begin{equation}  
\sum_{\kappa'}\left\{H^I_{\kappa\kappa'}-E^IN^I_{\kappa\kappa'}\right\}  
F^I_{\kappa'}=0.  
\label{psmeq}  
\end{equation}  
The expectation value of the Hamiltonian with respect to a ``rotational  
band $\kappa$'' $H^I_{\kappa\kappa}/N^I_{\kappa\kappa}$ defines    
a band energy, and when plotted as functions of spin $I$, we call  
it a band diagram \cite{review}.   
A band diagram displays bands   
of various configurations before they are mixed by the  
diagonalization procedure of Eq. (\ref{psmeq}).   
Irregularity in a spectrum may appear if a  
band is crossed by another one(s) at a certain spin.   
 
Fig. 1(a) shows our results for the   
lowest states of each spin, and at the prolate and oblate  
minima. They are presented in the plots of moment of inertia vs. square of  
rotational frequency. Remarkable agreement has been obtained for the oblate   
band. Our calculations suggest a gradual increase in the moment of inertia   
extended to higher spin states beyond the current experimental band.  
The basic features of the moment of inertia in the   
observed prolate band has also been reproduced.   
In addition to the sharp backbending at $I=8$   
that has been observed,   
we predict a second sharp backbending at $I=16$. We note that the  
highest spin state in the measurement of this band was $I=14$.   
Extension of the current experiment to higher spins can thus   
test our prediction.   
  
In Fig. 1(b), calculated B(E2) values are plotted for the prolate and oblate    
bands. These values have not been measured experimentally.   
We found that in the oblate band, the B(E2) values gradually increase   
until they are saturated at $I=14$.   
In the prolate band, however, drastic variations are predicted.  
Sudden drops in the curve appear at spins $I=8$ and $I=16$, corresponding  
to the two places where the moment of inertia bends back.    
From both the moment of inertia and B(E2) values,   
it is thus interesting to see that   
the co-existing bands in one nucleus exhibit very  
distinct characteristics where the prolate band is highly irregular and  
the oblate one is rather smooth.  
  
These interesting observations can be understood by studying the band diagrams.  
We found that the   
$g_{9/2}$ orbitals play a decisive role for the near yrast structure   
of this nucleus.  
Close to the neutron and proton Fermi levels of $^{68}$Se,   
single qp states of   
$g_{9/2}$ with different K quantum numbers   
are split by the deformed potential.    
At the prolate minimum, the Fermi levels are surrounded by the low-K states,  
whereas at the oblate minimum, they are near the high-K states.  
The low- and high-K states respond rather differently to the rotation,  
which is reflected in very different pair alignment processes.  
  
In Fig. 2, various  
configurations are distinguished by different types of lines. The  
symbols represent the yrast states obtained after the  
configuration mixing. There are more than 20 bands in each calculation,  
but only representative ones are displayed for discussion.  
Note that for the 2-qp bands, there are two closely-lying bands (a neutron   
and a proton band) because they nearly coincide with each other for the entire  
spin region.   
  
Fig. 2(a) shows the band diagram calculated at the oblate minimum.  
There are two 2-qp bands starting at about 5 MeV.    
These two 2-qp bands (one neutron and one proton  
band) are based on the $g_{9/2}$ quasiparticles with $K=7/2$ and $9/2$,   
coupled to $K=1$. As spin increases, they gradually approach the g-band,  
and finally cross with it between $I=14$ and 16. The band crossing is so gentle  
with a very small crossing angle that one cannot see a band disturbance   
in the yrast solution. In fact, rather smooth behavior has been predicted  
for the yrast band in both moment of inertia and B(E2) values (see Fig. 1).  
On the other hand, the two $g_{9/2}$ quasiparticles with $K=7/2$ and $9/2$  
can also couple to a 2-qp state with a total $K=8$.   
The $K=8$ bands have similar bandhead energies as the $K=1$ bands  
since they originate in the same two $g_{9/2}$ quasiparticles.   
  
Fig. 2(b) presents the band diagram at the prolate minimum.  
In this case, the two $g_{9/2}$ quasiparticles   
with low-K quantum numbers $K=1/2$ and $3/2$ lie close to the Fermi energy,   
and 2-qp bands involving these low-K states behave very differently  
from the oblate case.  
It can be seen that the two 2-qp bands (again, one neutron- and one   
proton-band) based on the low-K $g_{9/2}$ quasiparticles   
(coupled to $K=1$) sharply across the g-band at $I=8$.  
Because of this crossing, the g-band is disturbed, resulting in  
the observed first bandbending (see Fig. 1(a)).  
The crossing changes also the g-band dominance in the yrast wave functions   
before the crossing to the 2-qp band dominance after the crossing, which  
leads to the sudden drop in the B(E2) values (see Fig. 1(b)).  
  
A 4-qp  
state can be formed by coupling of  
the neutron and the proton 2-qp states discussed above.   
As shown by the dashed-dotted curve in Fig. 2(b), the 4-qp band can  
sharply across the 2-qp bands between $I=14$ and 16, and dominants the  
yrast wave functions thereafter.  
The crossing will result in the second backbending in moment of inertia,  
as shown in Fig. 1(a), and a bigger drop in B(E2) values at $I=16$, as  
predicted in Fig. 1(b).  
The highest spin state of the current experimental data   
for this band is $I=14$ \cite{Fi00}. Extension of a few high spin states  
will be a crucial test of our predictions, thus the correctness of the  
band crossing interpretation for $^{68}$Se.   
  
The theoretical energy spectrum at the oblate minimum is shown in Fig. 3   
in comparison with available data \cite{Fi00}.  
We predict two nearly-degenerate (only one of them is shown here)   
high-K bands ($K=8$) with a bandhead   
spin $I=8$ and an excitation energy of 5 MeV. They are   
long-lived isomeric states in the sense that no allowed $\gamma$-transition  
matrix elements of low multipolarity   
can connect these states to the nearby g-band ($K=0$).  
Our calculation shows the high-K  
band at a  lower excitation energy than any other 2-qp bands.  
Step-wise transitions via other multi-qp states  
are also unlikely since there are no other nearby states available.  
Thus, once the isomeric $I=8$ states are populated,  
the high-K states find no path for further $\gamma$-decay.   
  
Our prediction on the high-K isomeric states was based on a realistic   
calculation that has reasonably reproduced all the known structure data   
in this nucleus. However, we should mention two possible caveats.  
First, these isomers are based on 2-qp states  
of the $g_{9/2}$ orbitals. Therefore, correct single-particle energies of  
the neutron and proton $g_{9/2}$ orbitals are important.   
In the present calculation, we employed the newly-adjusted Nilsson  
parameters \cite{Sun00} which were obtained by a best estimation   
for the proton $g_{9/2}$ energy. Thus, further experimental data  
regarding an accurate proton $g_{9/2}$ position are very much desired.    
Second, neutron-proton pairing has not been explicitly   
considered in our theory.   
An interesting aspect in our case is that if the proton-neutron pairing   
interaction is included in our Eq. (\ref{hamham}),   
the two nearly-degenerate 2-qp states will no longer be proton or  
neutron 2-qp state, but a superposition of them. The interaction will  
generally push one of them down, and the other up, thus  
modifying their positions. Experimentally,   
if one would observe only one of them,  
that could be a signature of the proton-neutron pairing effect which  
may not be sensitively felt by the ground state.    
  
The occurrence of the $I=8$ isomers that lie   
very close to the yrast band may have   
a significant  
impact on the mass measurement for this nucleus, 
and a significant influence on the resulting 
abundances of the rp-process.  
The possibility that the GANIL measurement \cite{Lalle2000}  
has hit any isomeric state with an excitation greater than 1 MeV would 
explain the discrepancy with the fragmentation experiments.  
It should be noted also that the shape isomer, the lowest state 
in the prolate minimum, which has not been seen in Ref. \cite{Fi00},  
lies possibly at about 1 MeV excitation \cite{Moya99}.  
Thus, experimentally searching for these isomeric states is  
very important.   
 
In summary, we calculate the structure of $^{68}$Se using the PSM.  
Our results show excellent agreement with recent experimental  
findings regarding the spectroscopy of this nucleus:  
the coexisting oblate and prolate minima, the backbend at $I=8$.  
We further predict an additional backbend at $I=16$  
and a number of high K isomers at approximately 5 MeV above the ground state.  
The existence of isomeric states in this nucleus may provide resolution 
for the observed differences between  
the different types of measurements, and  
it may play a significant role in the mass  
flow and therefore resulting abundances of the rp-process.  
The latter point will be investigated further via network calculations  
of the rp-process.  
   
Valuable discussions with J.A. Sheikh   
and C.J. Lister are acknowledged.    
Y.S. thanks the colleagues at the   
University of Notre Dame  
for their warm hospitality, where this collaboration was initiated.  
The support of the National Science Foundation under grant PHY 99-01133  
and the graduate school of the University of Notre Dame are   
gratefully acknowledged.

\baselineskip = 14pt  
\bibliographystyle{unsrt}

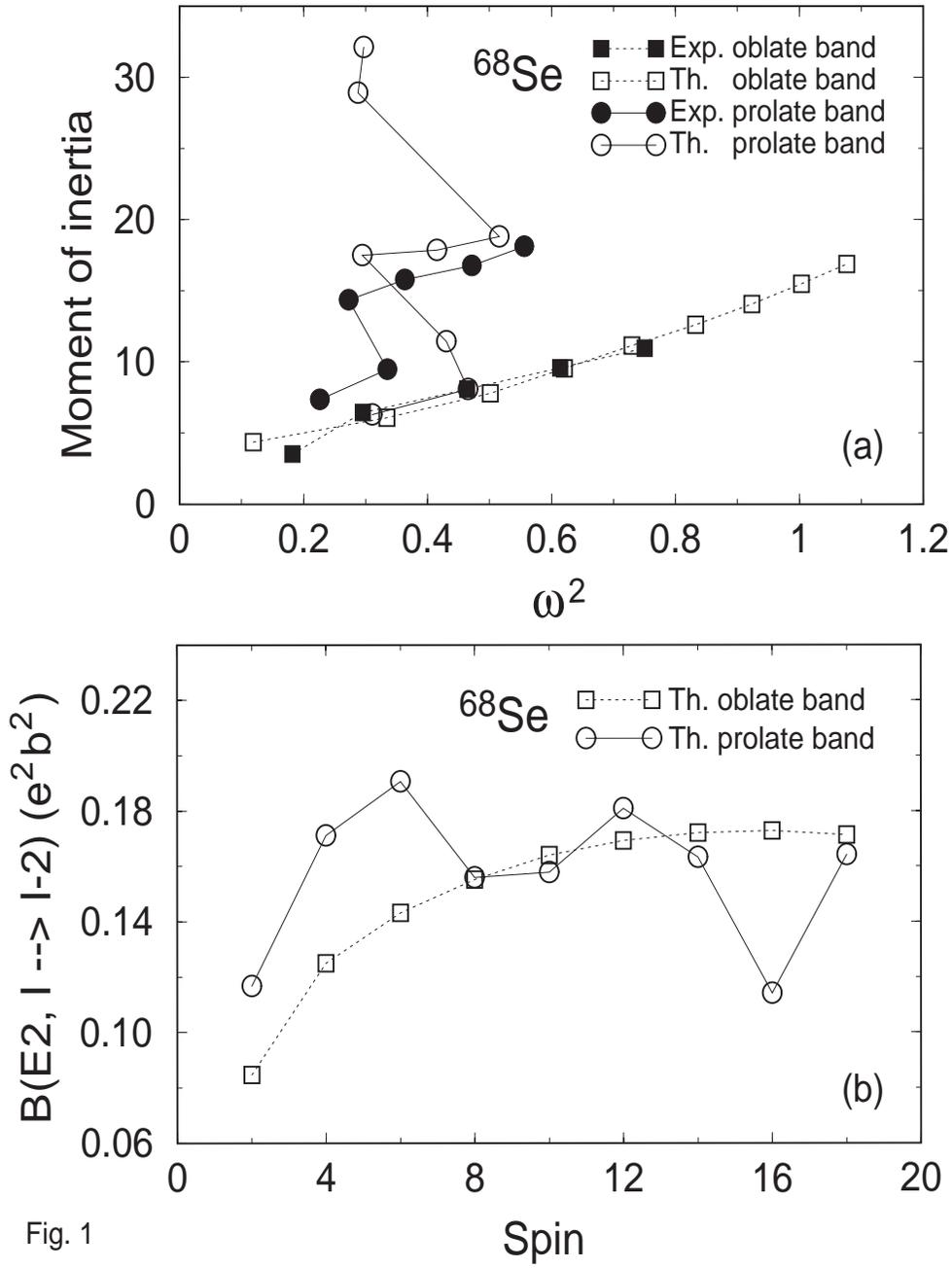
\begin{figure}  
\caption{(a) Moments of inertia ${2I-1}\over {E(I)-E(I-2)}$  
as function of $\omega^2$ with $\omega={{E(I)-E(I-2)}\over 2}$  
(The experimental data are taken from Ref.   
\protect\cite{Fi00}), and   
(b) calculated B(E2) values.   
}  
\label{figure.1}  
\end{figure}  
  
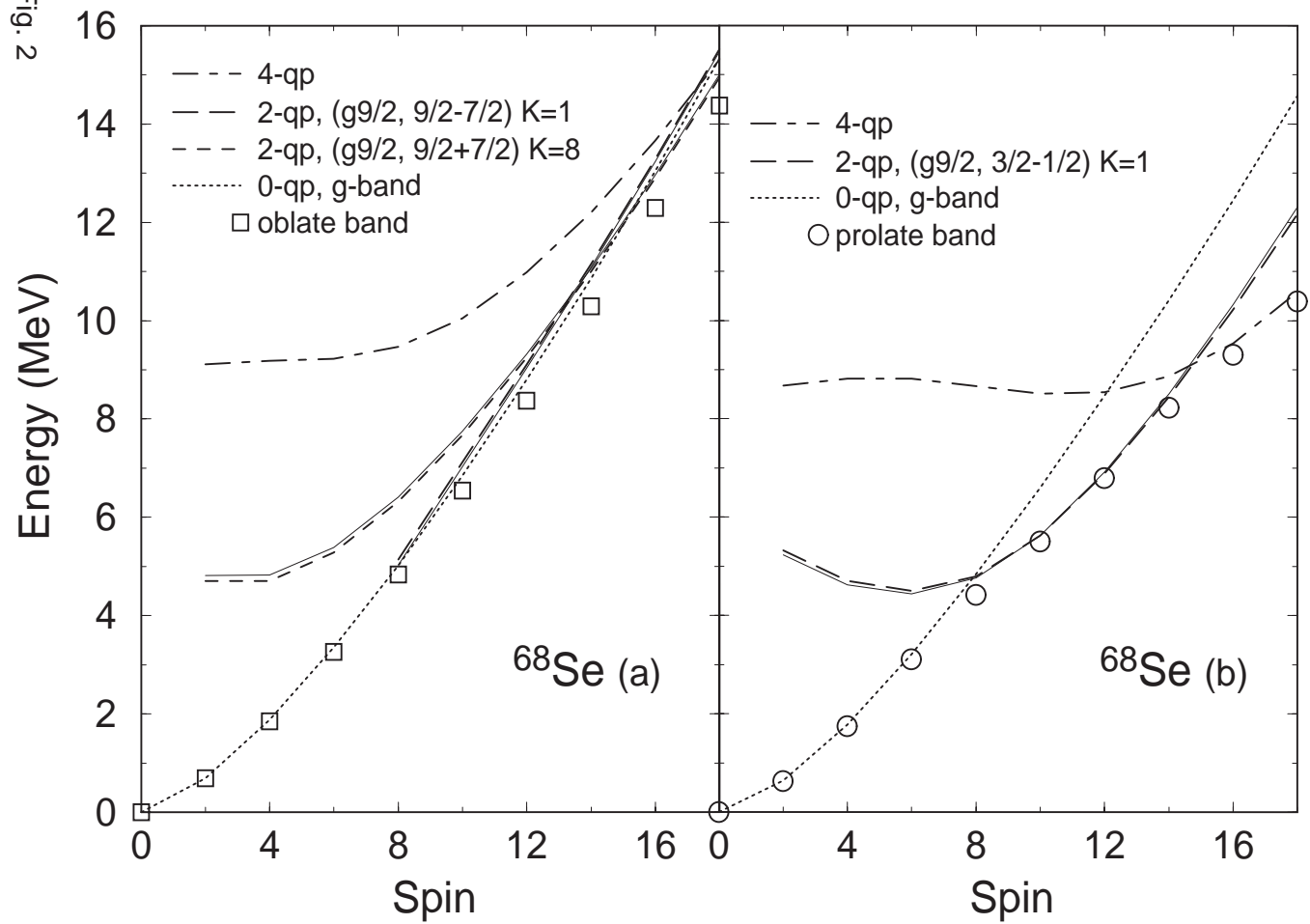
\begin{figure}  
\caption{  
Band diagrams (bands before configuration mixing) and 
the lowest band after configuration mixing (denoted by symbols)   
for $^{68}$Se at (a) the oblate minimum, and (b) the prolate minimum.   
Only the  
important lowest-lying  
bands in each configuration are shown.   
}  
\label{figure.2}  
\end{figure}  
  
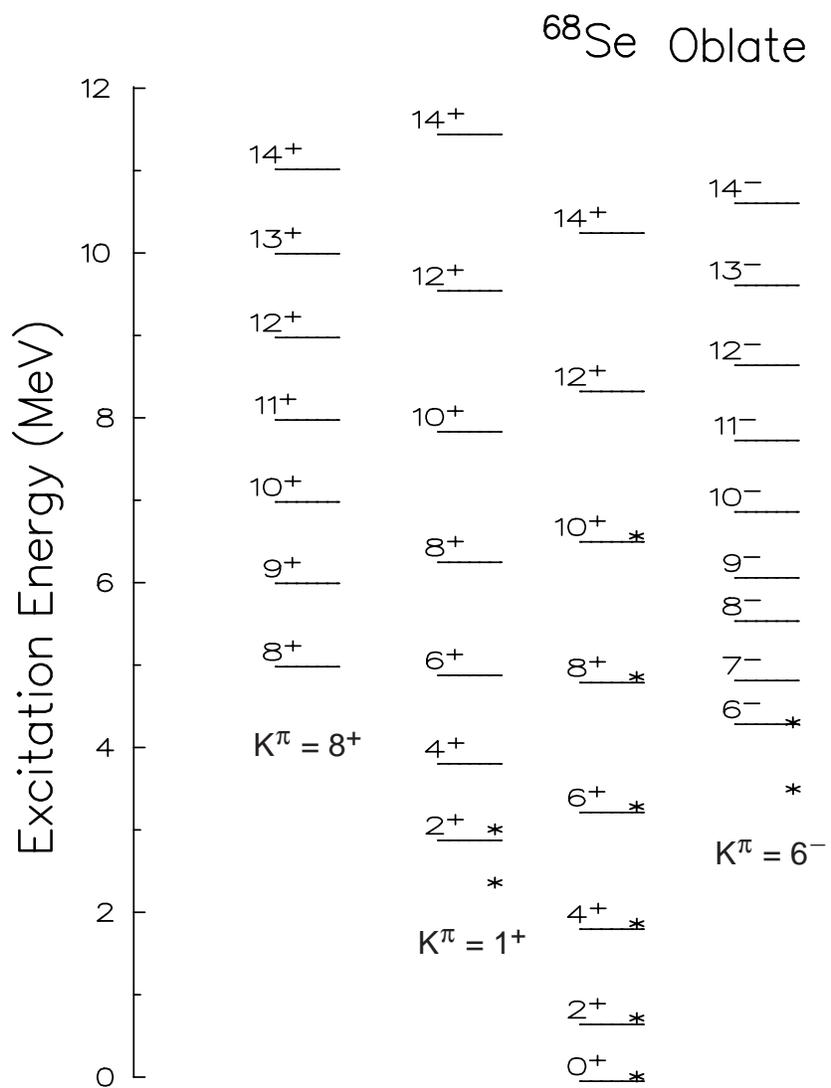
\begin{figure}  
\caption{Calculated energy spectrum for $^{68}$Se at the oblate minimum.  
High-K isomeric states are predicted. Results are compared with data  
(denoted by stars) whenever levels are experimentally known \protect\cite{Fi00}. 
}
\label{figure.3}  
\end{figure}  
  
\end{document}